\documentclass[conference]{IEEEtran}

\usepackage{cite}      

\usepackage[usenames,dvipsnames,svgnames,table]{xcolor}
\usepackage{graphicx}  

\usepackage{psfrag}    

\usepackage{url}       
\usepackage{tikz}

\usepackage{amssymb}
\usepackage{mathtools}

\usepackage{amsmath}   
\newcommand{\shorten}[1]{}

\newtheorem{example}{Example}

\begin{document}

\title{Coded Packet Transport for Optical Packet/Burst Switched Networks}

%
\author{\authorblockN{
Katina Kralevska, Harald {\O}verby and Danilo Gligoroski}
\authorblockA{Department of Telematics \\ Norwegian University of Science and Technology, Trondheim, Norway,\\ Email:
katinak@item.ntnu.no, haraldov@item.ntnu.no, danilog@item.ntnu.no}
}


\maketitle

\begin{abstract}
This paper presents the Coded Packet Transport (CPT) scheme, a novel transport mechanism for Optical Packet/Burst Switched (OPS/OBS) networks. The CPT scheme exploits the combined benefits of source coding by erasure codes and path diversity to provide efficient means for recovering from packet loss due to contentions and path failures, and to provide non-cryptographic secrecy. In the CPT scheme, erasure coding is employed at the OPS/OBS ingress node to form coded packets, which are transmitted on disjoint paths from the ingress node to an egress node in the network. The CPT scheme allows for a unified view of Quality of Service (QoS) in OPS/OBS networks by linking the interactions between survivability, performance and secrecy. We provide analytical models that illustrate how QoS aspects of CPT are affected by the number of disjoint paths, packet overhead and processing delay.

\end{abstract}


%
\IEEEpeerreviewmaketitle

\vspace{0.5cm}
\section{Introduction}
\vspace{0.1cm}
Optical Packet/Burst Switching (OPS/OBS) is a promising architecture for the future core network, enabling all-optical packet transport combined with statistical multiplexing for increased link utilization \cite{868150}. By avoiding electronic processing of packets, OPS/OBS achieves significant energy savings compared to existing opaque packet switched architectures \cite{Tucker:08}. The increasing number of mission-critical services such as e-banking, e-voting and emergency services put a high demand on the Quality of Service (QoS) of the future Internet, including OPS/OBS networks. Specifically, the OPS/OBS network has to provide low packet loss rate (performance) \cite{927342,Overby20071229,Overby2007,1197932}, protection against node and link failures (survivability) \cite{1242966,Overby:08}, as well as being able to withstand targeted eavesdropping attacks from individuals and organizations (secrecy) \cite{Medardbook}.

Existing approaches to satisfy these strict QoS demands in OPS/OBS rely on the provision of several independent QoS schemes, e.g., wavelength conversion to reduce packet loss from contentions and 1+1 path protection to provide survivability. However, since these schemes are deployed in the same physical and logical infrastructure, they will interact and provide mutual benefits. Examples of this include how the extra redundancy introduced for providing 1+1 path protection may be used to combat packet loss in OPS networks in failure-free operations, as studied in \cite{Overby:08}. In particular, security threats in all-optical networks have recently received research attention\cite{4597102,conf/ndss/MedardMC98}. One crucial security threat is eavesdropping of data in the network, which has traditionally been countered using encryption. However, the high capacities of OPS/OBS networks greater than 100 Gb/s make data encryption in OPS/OBS not feasible as the current computational resources do not match the required encryption processing demands. Hence, there is a need for a low complexity scheme that provides a certain level of secure data transport without encrypting the data. Our goal is to show how erasure coding and path diversity can be used to mutually provide loss recovery from contentions, survivability and a secrecy of data.

The major contribution of this paper is the novel Coded Packet Transport (CPT) scheme for OPS/OBS networks. This scheme is able to recover lost data due to contentions and node/link failures, while at the same time providing secrecy. We use the term secrecy as defined in \cite{Medardbook}, where the goal is protection from a passive adversary that is not able to reconstruct the whole packet/burst set by eavesdropping on a single path. The CPT scheme is based on Forward Error Correction (FEC) codes used as erasure codes and provides non-cryptographic secrecy. The CPT scheme is applicable to both OPS and OBS networks, and for the remainder of the paper we use the term packet to also refer to a burst in OBS networks, without the loss of generality. At an OPS/OBS ingress node, a set of data packets is encoded into a set of coded packets by utilizing non-systematic erasure codes \cite{1375250,Overby:04,1666647}. These coded packets are transmitted to an egress node in the OPS/OBS network on multiple disjoint paths. At an OPS/OBS egress node, reconstruction of packets lost due to contentions and node/link failures is enabled by the added redundancy. Sending different subsets of packets over disjoint paths between the ingress and the egress node also enables an end-to-end secrecy property against a passive adversary. To the best of our knowledge, this work constitutes a first step for providing a unified view on QoS in OPS/OBS networks, focusing on the interactions between survivability, performance and secrecy. 

The rest of this paper is organized as follows: Section \ref{Related} discusses related works. In Section \ref{CPT} we present the CPT scheme. Section \ref{Constraints} presents the analytical model. The parameter settings based on the analytical model are presented in Section \ref{Setting}. Finally, Section \ref{Conclusion} concludes the paper.

\section{Related works}\label{Related}
\vspace{0.05cm}

The authors of \cite{Overby:04} and \cite{1666647} show how FEC codes can be used to reduce packet loss in OPS networks. Here, redundant packets are added to a set of data packets at the OPS ingress node and transmitted along with the original data packets to an OPS egress node. Data packets dropped due to contentions may be reconstructed at the OPS egress node by using excess redundant packets, leading to a potential reduced Packet Loss Rate (PLR). The work in \cite{Overby:08} and \cite{overby2012icc} extends these schemes to provide 1+1 path protection. One redundant packet is added to a packet set using the XOR operation. This packet set is transmitted to the OPS egress node over three or more disjoint paths. In particular, the authors of \cite{overby2012icc} evaluate this scheme from a cost perspective, comparing it to other approaches that provide 1+1 path protection, showing that significant cost savings can be achieved using erasure codes. Unlike the present paper, these schemes do not consider secrecy.

The authors of \cite{Nguyen02distributedvideo} show how the PLR can be significantly reduced by sending packets at appropriate rates on disjoint paths from multiple ingress nodes to an egress node by using FEC techniques.
The work is further extended in \cite{1208716} where a scalable, heuristic scheme for selecting a redundant path between an ingress node and an egress node is presented. Another way of reducing the packet loss due to contentions in OPS networks is by combining source and network coding. Instead of dropping the colliding packets at the intermediate node, they are XOR-ed together \cite{Gergely2011,7034478}. 

The authors of \cite{4597102} suggest an OBS framework that provides authentication of burst headers and confidentiality of data bursts based on encryption. However, due to the high bandwidth in OBS networks, the encryption mechanisms have to be with low computational complexity, suitable for high-speed implementation and the majority of the header content should not be encrypted since the processing of the headers has to be at ultra high speed \cite{4597102}. Unlike their work, we provide a certain level of non-cryptographic secrecy in the data transport without using encryption, thus significantly reducing the computational complexity. 

That secrecy is achieved by sending non-systematic coded packets on disjoint paths. This property has so far not been exploited in OPS/OBS networks. The authors of \cite{journals/tifs/OliveiraLVBM12} show how to provide secrecy in storage systems even when an eavesdropper knows or can guess some of the missing symbols. This is achieved by using MDS matrices (matrices that have no singular square submatrices). On the other hand, schemes based on non-systematic codes increase the delay in the networks, as decoding of non-systematic coded packets can start after the number of received coded packets is at least equal to the number of original data packets. In addition, the encoding and decoding of the packets also add processing delay in the network.

\vspace{0.5cm}
\section{Coded Packet Transport (CPT)}\label{CPT}

\begin{figure}
\centering
\includegraphics[width=3.5in]{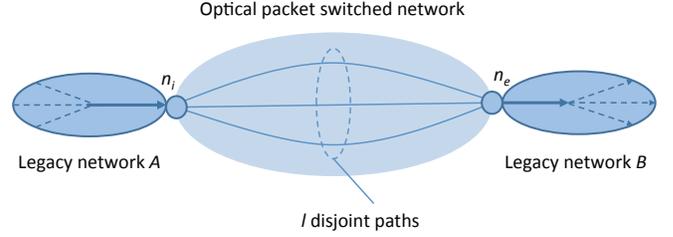}
\caption{An OPS network where data is transmitted from ingress node $n_i$ to egress node $n_e$}
\label{fig_net}
\end{figure}

\begin{figure}
\centering
\includegraphics[width=3.5in]{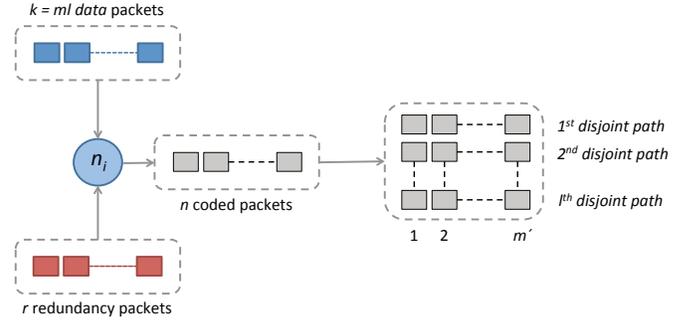}
\caption{Illustration of CPT, $k$ data packets are encoded into $n$ coded packets at $n_i$ and transmitted over $l$ disjoint paths}
\label{fig_CPT}
\end{figure}

We focus on an ingress and egress node pair ($n_i$, $n_e$) in an OPS/OBS network as depicted in Fig.  \ref{fig_net}. Packets arrive at $n_i$ from a legacy network A, with destination $n_e$ or a legacy network B. We assume that there exist $l$ disjoint paths between $n_i$ and $n_e$. We use the term disjoint paths to refer to node disjoint (respectively link disjoint) paths between two nodes in the network. The ingress node $n_i$ encodes the $k$ data packets into $n$ coded packets. The disjoint paths are independent and disjunctive subsets of the coded packets are sent on them. Figure \ref{fig_CPT} depicts the encoding of $k$ data packets with equal length into $n$ coded packets. Since we analyze the effect of $l$ on the QoS in OPS/OBS, both the original and coded packets are written as multiple of $l$, i.e., $k=m l$ and $n= m' l$. Here we introduce the metric packet overhead $o$ that is a ratio of the number of redundant packets and data packets, i.e., $o=\frac{r}{k}$. Notations used in the paper are summarized in Table \ref{SporedbaSite}. Packets in the network may be lost due to contention inside packet switches and due to node/link failures \cite{Overby:08}.
Note that we exclude the ingress and egress node from the set of nodes that can fail.

The CPT scheme is based on FEC codes that are used as erasure codes. Different coding schemes may be applied for the CPT, including Maximum Distance Separable (MDS) and non-MDS codes. Let us denote by $GF(2^q)$ the Galois field with $2^q$ elements. An $(n, k, d)_{2^q}$ code is an $(n, k)_{2^q}$ code of length $n$ and rank $k$ with minimum weight $d$ among all nonzero codewords. An $(n, k, d)_{2^q}$ code is called MDS if $d = n - k + 1$. The Singleton defect of an $(n, k, d)_{2^q}$ code $C$ defined as $s(C ) = n - k + 1 - d$ measures how far away is $C$ from being MDS.
Reed-Solomon codes are the most commonly used MDS codes, i.e., their Singleton defect is zero \cite{Reed:1960:PCC}. When Reed-Solomon codes are used, at least $k$ out of $n$ packets must arrive successfully at the egress node in order to enable recovery of the data.

An $(n, k)$ linear block code is defined by its $n \times k$ generator matrix $\mathbf{G}$. A code is systematic if the first $k$ rows of its generator matrix $\mathbf{G}$ contain the identity matrix. That means that a systematic linear code does not transform the original $k$ data packets, but generates only extra $n-k$ redundant packets.

If the generator matrix $\mathbf{G}$ is not systematic then the code is non-systematic, and all $n$ generated packets linearly depend on all original $k$ packets via $\mathbf{G}$. Systematic codes are less processing demanding than non-systematic, since they do not require processing of the original data. If the goal is to achieve a certain level of secrecy in the transmitted data, then non-systematic codes should be used. That is the main reason why we use non-systematic Reed-Solomon codes.

An $(n, k)$ Reed-Solomon code is obtained by evaluating polynomials over $GF(2^q)$ at $n$ different points $\alpha_1, \alpha_2, \ldots, \alpha_n$. The generator matrix for Reed-Solomon code is an $n \times k$ Vandermonde matrix

\begin{equation*}
\mathbf{G}=
\begin{bmatrix}
1 & \alpha_1 & {\alpha_1}^2 & \ldots & {\alpha_1}^{k-1}\\
1 & \alpha_2 & {\alpha_2}^2 & \ldots & {\alpha_2}^{k-1}\\
\vdotswithin{1} & \vdotswithin{\alpha_n} & \vdotswithin{{\alpha_n}^2} & \ddots & \vdotswithin{{\alpha_n}^{k-1}}\\
1 & \alpha_{n-1} & {\alpha_{n-1}}^2 & \ldots & {\alpha_{n-1}}^{k-1}\\
1 & \alpha_n & {\alpha_n}^2 & \ldots & {\alpha_n}^{k-1}\\
\end{bmatrix}.
\end{equation*}

The encoding in CPT is performed in the following way.

\begin{table}[t]
\caption{Overview of parameters}
\centering
\begin{tabular}{*{15}{c}}
\hline
$n_i$ & Ingress node\\
\hline
$n_e$ & Egress node\\
\hline
$k$ & Number of original data packets\\
\hline
$n$ & Number of coded packets\\
\hline
$r$ & Number of redundant packets\\
\hline
$L$ & Packet length\\
\hline
$l$ & Number of disjoint paths\\
\hline
$m$ & Number of data packets sent on a disjoint path\\
\hline
$m'$ & Number of coded packets sent on a disjoint path\\
\hline
$GF(2^q)$ & Galois field\\
\hline
$p$ & Packet Loss Rate\\
\hline
$p_{thres}$ & Packet Loss Rate threshold \\
\hline
$o$ & Packet overhead\\
\hline
\end{tabular}
\label{SporedbaSite}
\end{table}

We assume that the $k$ original data packets with equal length consist of $L$ symbols from $GF(2^q)$, i.e., $x_i=(s_{i, j})$ where $1 \leq i \leq k$ and $1 \leq j \leq L$. The $k$ data packets are presented in a form of a matrix $\mathbf{X}$ where every row represents one data packet

\begin{equation*}
\mathbf{X}=
\begin{bmatrix}
s_{1, 1} & s_{1, 2}  & \ldots & s_{1, L}\\
s_{2, 1} & s_{2, 2}  & \ldots & s_{2, L}\\
\vdotswithin{s_{k_1}} & \vdotswithin{s_{k_2}} & \ddots & \vdotswithin{{\alpha_n}^{k-1}}\\
s_{k, 1} & s_{k, 2} & \ldots & s_{k, L}\\
\end{bmatrix}.
\end{equation*}

We obtain the coded packets by simple matrix multiplication of $\mathbf{G}$ with $\mathbf{X}$, i.e.,

\begin{equation*}
\mathbf{Y}=\mathbf{G} \times \mathbf{X}.
\end{equation*}

Similarly as in $\mathbf{X}$, every row of $\mathbf{Y}$ represents an encoded packet.
For the sake of clarity we represent the matrix $\mathbf{Y}$ in $l$ stripes of $m'=\lceil \frac{n}{l} \rceil$ rows, since we consider transmitting data on $l$ disjoint paths

\begin{equation*}
\mathbf{Y}=
\begin{bmatrix}
s_{1, 1} & s_{1, 2}  & \ldots & s_{1, L}\\
\vdotswithin{s_{k_1}} & \vdotswithin{s_{k_2}} & \ddots & \vdotswithin{{\alpha_n}^{k-1}}\\
s_{m', 1} & s_{m', 2}  & \ldots & s_{m', L}\\
 \hline
s_{m'+1, 1} & s_{m'+1, 2}  & \ldots & s_{m'+1, L}\\
\vdotswithin{s_{k_1}} & \vdotswithin{s_{k_2}} & \ddots & \vdotswithin{{\alpha_n}^{k-1}}\\
s_{2m', 1} & s_{2m', 2}  & \ldots & s_{2m', L}\\
 \hline
 \vdotswithin{s_{k_1}} & \vdotswithin{s_{k_2}} & \ddots & \vdotswithin{{\alpha_n}^{k-1}}\\
 \hline
s_{{(l-1)m'+1},1} & s_{{(l-1)m'+1}, 2}  & \ldots & s_{{(l-1)m'+1}, L}\\
\vdotswithin{s_{k_1}} & \vdotswithin{s_{k_2}} & \ddots & \vdotswithin{{\alpha_n}^{k-1}}\\
s_{n, 1} & s_{n, 2} & \ldots & s_{n, L}\\
\end{bmatrix}.
\end{equation*}

The packets from the $i-$th stripe are sent on the $i-$th disjoint path, respectively. Once again we emphasise that sending linear combinations of the data packets instead of using a systematic code offers secrecy as opposed to systematic codes.

Next we discuss the required parameter settings for enabling survivability, secrecy and performance within a predefined PLR threshold.

\vspace{0.3cm}
\section{Analytical model}\label{Constraints}

\subsection{Topology constraints}
\vspace{0.05cm}
The parameter $l$ dictates the number of disjoint paths between a pair of nodes in the network. We choose to vary $l$ between 2 and 6, which is grounded in the constraints provided by most empirical network topologies. Hence, $1+N$ protection for a big $N$ is impractical. In the following analysis we combine path diversity with source coding to meet the goals of the CPT.

\subsection{Survivability constraints}
In order to provide 1+1 path protection against single link failure, the number of received packets at $n_e$ must be equal to or larger than the original number of data packets, $k$. Hence, as the number of lost data packets in the case of a failed path is $m'$, we have that

\begin{equation}
n - m' \geq k,
\label{rhos}
\end{equation}

resulting in the following constraint for $o$

\begin{equation}
o \geq \frac{1}{l-1}.
\label{survivability}
\end{equation}

\vspace{0.3cm}
\subsection{Secrecy constraints}
\vspace{0.05cm}
The goal is to achieve secrecy so that a passive adversary is not able to reconstruct the whole packet set by eavesdropping on a single path. We use here the term secrecy as it is used in \cite[Ch.7 pp. 185]{Medardbook}.

An eavesdropper needs to eavesdrop $k$ packets in order to decode the whole packet set. By eavesdropping on a single path, $m'$ packets are obtained. Hence, we ensure that by eavesdropping on a single path it is not possible to recover the data packets if the following

\begin{equation}
k \geq m'
\label{rhos}
\end{equation}
 is fullfiled. Resulting in the following constraint for $o$

\begin{equation}
o \leq l-1.
\label{secrecy}
\end{equation}

One straightforward attack on the secrecy of CPT is by applying the strategy described by McEliece and Sarwate in \cite{mceliece1981sharing}. The authors conclude that decoding algorithms for Reed-Solomon codes provide extensions and generalizations of the Shamir's secret sharing scheme \cite{shamir1979share}.

We adapt the strategy discussed in the last paragraph of \cite{mceliece1981sharing}. Namely, the attack is a combination of a partially known coded text and a brute force attack. An eavesdropper knows or guesses the format or even the content of some parts of the coded information and performs an exhaustive search to check all possible linear combinations in order to filter out the wrong guesses and to find the correct ones. We give a short example where we show that constraint \ref{secrecy} is not enough for providing secrecy when an eavesdropper is able to guess the missing packets, i.e., run a brute force attack.
\begin{example}
Based on constraints \ref{survivability} and \ref{secrecy}, the network is in the operational range of CPT if $l=3$ and $o=1$. We define a range as operational when the secrecy and the survivability constraints are not violated. Let us take the following Reed-Solomon code (12, 6) over $GF(2^8)$. If we want to achieve survivability and secrecy, 4 packets are sent on each disjoint path. However, if an eavesdropper gets the coded packets from a single path, he/she gets 4 coded packets. The eavesdropper requires 2 more coded packets to decode the whole data set. Since the coding is performed over $GF(2^8)$, it applies an exhaustive search of $2^{2 \times 8} = 2^{16}$ tries for every 2 missing bytes from the 2 missing packets. Knowing the format of the submitted information (which is a reasonable and plausible assumption) the eavesdropper filters out all solutions that do not fit within its filtering strategy and keeps the solutions that give decoded information that has the expected format.
\end{example}

We combine the attack technique discussed above with the modern recommended levels of security as they are given in
\cite{BlueKrypt2014,lenstra2001selecting,barker2006recommendation}. The minimum level of security for 2015 is between 80 and 112 bits (Lenstra/Verhul, ECRYPT II, NIST), but we put a level of 128 bits as a long term security level. So we derive an additional constraint for secrecy in CPT considering these recommendations, i.e.,
\begin{equation}
(k-m') q \geq 128
\label{rhos}
\end{equation}

resulting in the following constraint for $o$
\begin{equation}
o \leq l-1 - \frac{128}{q m}.
\label{strong}
\end{equation}

\begin{example}
In this example we still assume that $l$=3 and $o$=1. The coding is performed with Reed-Solomon over $GF(2^8)$, but $n$ and $k$ are choosen following constraint \ref{strong}.
Let us take the Reed-Solomon code (96, 48) over $GF(2^8)$. On each disjoint path 32 packets are sent. If an eavesdropper gets the coded packets on a single path, still it misses 16 packets to decode the original data. Running an exhaustive search of $2^{16 \times 8} = 2^{128}$ tries for every 16 missing bytes from the 16 missing packets is infeasible.
\end{example}

\vspace{0.15cm}
\subsection{Performance constraints}
When the PLR is $p$ and the loss of the redundant packets is accounted, then the average number of redundant packets is
\begin{equation}
r = \frac{k p}{1-p}.
\label{rhos}
\end{equation}

However, adding $r$ redundant packets does not guarantee that the PLR is kept bellow a predefined threshold $p_{thres}$.

Unsuccessful decoding occurs when more than $r$ packets are lost. The probability of unsuccessful decoding $p_{fail}$ when RS codes are used is
\begin{equation}
p_{fail} = \sum_{i=n-k+1}^{n} \binom{n}{i} p^{i}(1-p)^{n-i}.
\label{failure2}
\end{equation}

If one path has failed, still data packets on the nonfailed paths can be lost due to contentions. The required  number of redundant packets is calculated for $n'=n-m'$ and $p_{fail}\leq p_{thres}$, i.e., from the following expression
\begin{equation}
\sum_{i=n'-k+1}^{n'} \binom{n'}{i} p^{i}(1-p)^{n'-i} \leq p_{thres}.
\label{coding3}
\end{equation}

\vspace{0.15cm}
\subsection{Processing constraints}
Since in CPT we add redundancy of $n-k$ packets and we use non-systematic RS codes, we introduce processing delay. The processing delay is expressed in number of clock cycles.

The processing delay should not be confused with latency, because it is a component of the latency. Since decoding is more computationally demanding than encoding, we calculate the processing delay at the egress node. The processing delay is defined as the number of clock cycles required for a packet set to enter, be processed and leave the egress node.

In \cite{xilinx} the processing delay in number of clock cycles per symbol for a RS systematic code $(n, k)$ at the decoder  is calculated as
\begin{equation}
d_{proc}= (n-k)^2+6(n-k)+4.
\label{rhos}
\end{equation}

If the processing delay for the packet set is greater than the size of the packet set, $d_{proc} > n$, then the node should buffer the packets arriving from the next packet set while processing the previous one.

The time to process the packet set for a systematic code is dependent on the number of redundant packets $n-k$. For a non-systematic code, the size of the whole packet set $n$ should be counted, not just $n-k$. First we obtain the delay for processing a single packet and then we multiply it with the total number of packets in the packet set. Therefore, the processing delay in number of clock cycles per symbol for a RS non-systematic code $(n, k)$ at the decoder is
\begin{equation}
d_{proc}= \lceil \frac{n}{n-k} \rceil ((n-k)^2+6(n-k)+4).
\label{rhos}
\end{equation}

\vspace{0.35cm}
\section{Parameter settings for CPT}\label{Setting}
\vspace{0.1cm}

\begin{figure}
\centering
\includegraphics[width=3.5in]{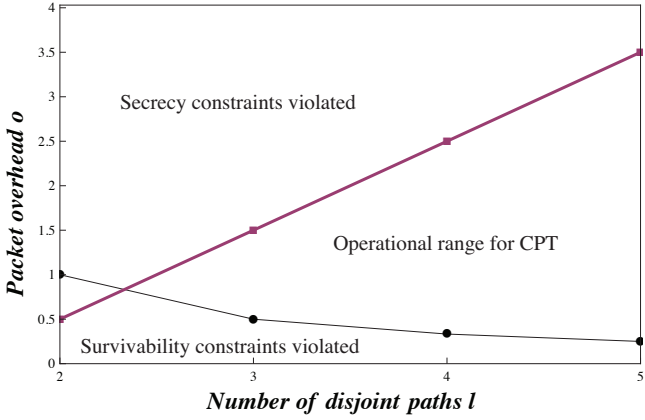}
\caption{The operational range for CPT for $m=32$ and coding over $GF(2^8)$}
\label{cpt}
\end{figure}

Figure \ref{cpt} shows the operational range of the CPT, i.e., the relationship between the number of available disjoint paths between a node pair ($l$) and the packet overhead ($o$). Three different areas are identified: \begin{enumerate}
\item The operational range for the CPT scheme;
\item The secrecy constraints are violated;
\item The survivability constraints are violated.
\end{enumerate}
As a general insight, we observe that a certain amount of packet overhead needs to be provided in order to ensure survivability (protection against one failure). Furthermore, sending a certain amount of data packets less than $k$ on each path should guarantee secrecy against eavesdropping. Thus, the combination of $o$ and $l$ should be in the operational range.  Figure \ref{cpt} is produced for $m=32$ and coding over $GF(2^8)$.  

\begin{figure}
\centering
\includegraphics[width=3.5in]{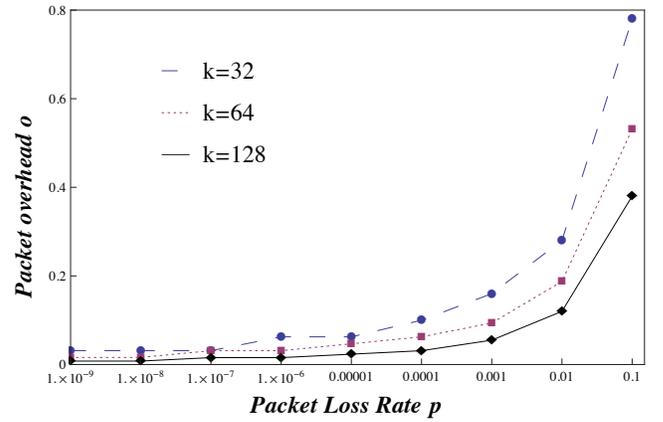}
\caption{Required packet overhead for different PLR so that $p_{fail} \leq p_{thres}$ for $p_{thres}$=$10^{-12}$}
\label{plr}
\end{figure}

Figure \ref{plr} depicts the required overhead ratio for different PLR when $k=32, 64$ and $128$ and $p_{thres}$=$10^{-12}$. Here it is shown that the packet overhead required to keep the PLR below a predefined $p_{thres}$ is less than the required packet overhead to achieve survivability against single link failure. 

\begin{figure}
\centering
\includegraphics[width=3.5in]{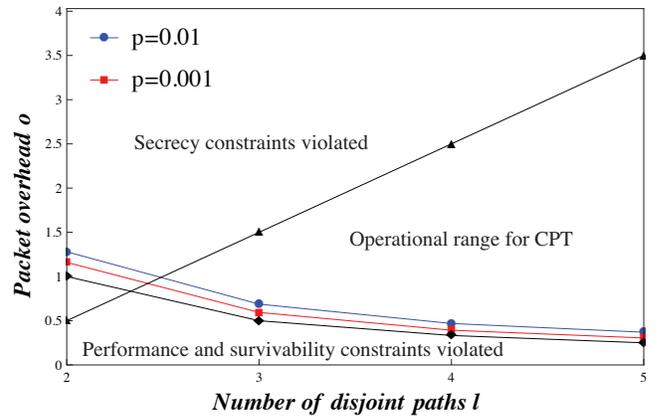}
\caption{The reduced operational range for CPT for $m$=$32$, coding over $GF(2^8)$ when $p$=$0.01$ and $0.001$ and $p_{thres}$=$10^{-12}$}
\label{Reducedcpt}
\end{figure}

Figure \ref{Reducedcpt} illustrates the operational range when we consider constraint (\ref{coding3}) in addition to the constraints for survivability and secrecy. The $n-m'$ packets sent on the nonfailed paths may still be lost due to contentions that means additional overhead is needed. The operational range reduces as the PLR increases. We draw the operational range for $p$=$0.01$ and $0.001$ besides the one presented in Figure \ref{cpt}.

In the previous Section we choose small parameteres for $n$ and $k$ for simplified explanation. We give other examples in Table \ref{param} where $n$ is equal to $2^q-1$ and the parameters are chosen to satisfy the survivavility and secrecy constraints.

\begin{table}[t]
\caption{Examples where $n=2^q-1$}
\centering
\begin{tabular}{*{15}{c}}
\hline
$q$ & $k$ & $r$ & $o$ & $m'$ & $l$ & Secrecy level in bits\\
\hline
$5$ & $25$ & $6$ & $0.24$ & $5$ and $6$ & $6$ & $95$\\
\hline
$6$ & $42$ & $21$ & $0.5$ & $21$ & $3$ & $126$\\
\hline
$7$ & $84$ & $43$ & $0.512$ & $42$ and $43$ & $3$ & $287$\\
\hline
$8$ & $204$ & $51$ & $0.25$ & $51$ & $5$ & $1224$\\
\hline
\end{tabular}
\vspace{0.5cm}
\label{param}
\end{table}

\begin{figure}
\centering
\includegraphics[width=3.5in]{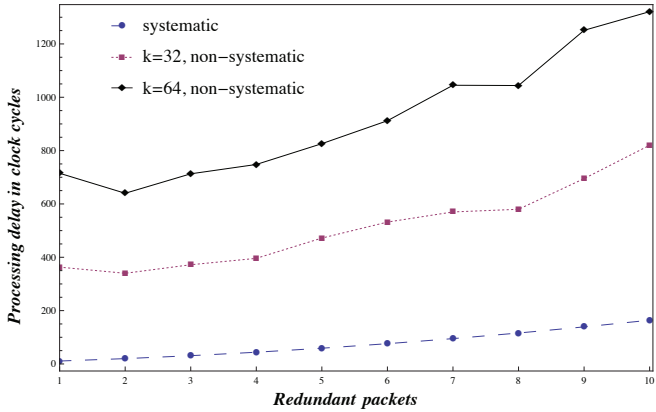}
\caption{Processing delay in clock cycles per symbol for a systematic and non-systematic RS code}
\label{delay}
\end{figure}

Figure \ref{delay} shows the processing delay for a systematic code and two non-systematic RS codes for $k=32$ and $64$. The processing delay for the non-systematic codes is significantly higher than for the systematic code. This is because that the total number of packets processed by a non-systematic code is greater than the number of redundant packets processed by a systematic code. The processing delay increases as $r$ increases for the both cases. As it is illustrated in Figure \ref{delay}, the number of data packets $k$ has an impact on the processing delay in addition to $r$ for a non-systematic code. When non-systematic coding is performed, the decoder will always process the packets from the previous packet set while packets from a new packet set arrive. This is not the case for a systematic code for some parameter selections.

We conclude that the survivability, performance and secrecy that the CPT offers are at a price of an increased processing delay. The CPT offers better performance compared to 1+1 or 1+N protection, where the overhead ratio is 1. In order to keep the CPT scheme in the operational range, either the overhead ratio should be greater than 0.5 or the number of disjoint paths should be equal to or more than 3. The CPT scheme offers both survivability and secrecy with less overhead compared to traditional 1+1 path protection, which offers only survivability against a single failure for 100$\%$ overhead ratio. 



\vspace{0.3cm}
\section{Conclusions}\label{Conclusion}

This paper presented the Coded Packet Transport (CPT) scheme, a novel transport mechanism for Optical Packet/Burst Switched (OPS/OBS) networks. The CPT scheme is able to recover packets lost due to contentions and node/link failures, as well as providing secrecy in OPS/OBS networks. We have presented the conceptual architecture for the CPT scheme, along with analytical results, outlining the achievable performance. Further research on this topic should investigate the performance in realistic network topologies.

\vspace{0.3cm}
\section*{Acknowledgements}\label{acknowledgemets}

We would like to thank Gergely Bicz\'{o}k for his discussions and remarks that significantly improved the paper.

\vspace{0.5cm}
\bibliographystyle{plain}

\bibliography{refer}
\vspace{0.2cm}

\end{document}